\documentstyle[11pt,aaspp4,epsfig]{article}
\begin{document}

% Peterson's LaTeX definitions

%

\def\IUE{{\it IUE}}
\def\HST{{\it HST}}
\def\Kronos{{\it Kronos}}
\def\bfKronos{\boldmath $K\!ronos$}
\def\deg{\ifmmode ^{\rm o} \else $^{\rm o}$\fi}
\def\degC{$^{\rm o}$C}
\def\arcsec{\ifmmode '' \else $''$\fi}
\def\arcmin{\ifmmode ' \else $'$\fi}
\def\arcsecpoint{\ifmmode ''\!. \else $''\!.$\fi}
\def\arcminpoint{\ifmmode '\!. \else $'\!.$\fi}
\def\kms{\ifmmode {\rm km\,s}^{-1} \else km\,s$^{-1}$\fi}
\def\Msun{\ifmmode M_{\odot} \else $M_{\odot}$\fi}
\def\Lsun{\ifmmode L_{\odot} \else $L_{\odot}$\fi}
\def\qo{\ifmmode q_{\rm o} \else $q_{\rm o}$\fi}
\def\Ho{\ifmmode H_{\rm o} \else $H_{\rm o}$\fi}
\def\ho{\ifmmode h_{\rm o} \else $h_{\rm o}$\fi}
\def\ltsim{\raisebox{-.5ex}{$\;\stackrel{<}{\sim}\;$}}
\def\gtsim{\raisebox{-.5ex}{$\;\stackrel{>}{\sim}\;$}}

\def\cc{\ifmmode {\rm cm}^{-3} \else cm$^{-3}$\fi}
\def\cm2{\ifmmode {\rm cm}^{-2} \else cm$^{-2}$\fi}
\def\CCF{\ifmmode F_{\it CCF} \else $F_{\it CCF}$\fi}
\def\ACF{\ifmmode F_{\it ACF} \else $F_{\it ACF}$\fi}
\def\Halpha{\ifmmode {\rm H}\alpha \else H$\alpha$\fi}
\def\Hbeta{\ifmmode {\rm H}\beta \else H$\beta$\fi}
\def\Hgamma{\ifmmode {\rm H}\gamma \else H$\gamma$\fi}
\def\Hdelta{\ifmmode {\rm H}\delta \else H$\delta$\fi}
\def\Lya{\ifmmode {\rm Ly}\alpha \else Ly$\alpha$\fi}
\def\Lyb{\ifmmode {\rm Ly}\beta \else Ly$\beta$\fi}
\def\hi{H\,{\sc i}}
\def\hii{H\,{\sc ii}}
\def\hei{He\,{\sc i}}
\def\heii{He\,{\sc ii}}
\def\ci{C\,{\sc i}}
\def\cii{C\,{\sc ii}}
\def\ciii{\ifmmode {\rm C}\,{\sc iii} \else C\,{\sc iii}\fi}
\def\civ{\ifmmode {\rm C}\,{\sc iv} \else C\,{\sc iv}\fi}
\def\ni{N\,{\sc i}}
\def\nii{[N\,{\sc ii}]}
\def\niii{N\,{\sc iii}}
\def\niv{N\,{\sc iv}}
\def\nv{N\,{\sc v}}
\def\oi{[O\,{\sc i}]}
\def\oii{O\,{\sc ii}}
\def\oiii{[O\,{\sc iii}]}
\def\o5007{[O\,{\sc iii}]\,$\lambda5007$}
\def\oiv{O\,{\sc iv}}
\def\ov{O\,{\sc v}}
\def\ovi{O\,{\sc vi}}
\def\nev{Ne\,{\sc v}}
\def\mgi{Mg\,{\sc i}}
\def\mgii{Mg\,{\sc ii}}
\def\siIV{Si\,{\sc iv}}
\def\si{S\,{\sc i}}
\def\sii{[S\,{\sc ii}]}
\def\siii{S\,{\sc iii}}
\def\caii{Ca\,{\sc ii}}
\def\feii{Fe\,{\sc ii}}
\def\fevii{Fe\,{\sc vii}}
\def\fex{Fe\,{\sc x}}
\def\aliii{Al\,{\sc iii}}
\def\fun#1#2{\lower3.6pt\vbox{\baselineskip0pt\lineskip.9pt
  \ialign{$\mathsurround=0pt#1\hfil##\hfil$\crcr#2\crcr\sim\crcr}}}
\def\lap{\mathrel{\mathpalette\fun <}}
\def\gap{\mathrel{\mathpalette\fun >}}
\def\lax    {${_<\atop^{\sim}}$}
\def\gax    {${_>\atop^{\sim}}$}
\def\MBH{\ifmmode M_{\bullet} \else $M_{\bullet}$\fi}
\def\Msigma{\ifmmode M_{\bullet}\,\mbox{--}\,\sigma 
	\else $M_{\bullet}$ -- $\sigma$\fi}

\newcommand{\vFWHM}{\mbox{$V_{\mbox{\tiny FWHM}}$}}
\newcommand{\Tdur}{\mbox{$T_{\rm dur}$}}
\newcommand{\Tres}{\mbox{$\Delta t$}}
\newcommand{\vres}{\mbox{$\Delta v$}}
\newcommand{\Rin}{\mbox{$R_{\rm in}$}}
\newcommand{\Rout}{\mbox{$R_{\rm out}$}}

\title{Observational Requirements for High-Fidelity Reverberation Mapping}

\author{
	Keith~Horne,\altaffilmark{1,2}
	Bradley~M.~Peterson,\altaffilmark{3}
	Stefan~J.~Collier,\altaffilmark{3}
	and Hagai~Netzer\,\altaffilmark{4}
	}
\altaffiltext{1}{
	School of Physics and Astronomy,
	University of St. Andrews, Fife KY16~9SS, Scotland, UK
\\	kdh1@st-and.ac.uk
	}
\altaffiltext{2}{
	Department of Astronomy,
	University of Texas, Austin, TX 78712
	}
\altaffiltext{3}{
	Department of Astronomy, The Ohio State University, 
	140 West 18th Avenue, Columbus, OH 43210
\\	peterson@astronomy.ohio-state.edu
}
\altaffiltext{4}{
	School of Physics and Astronomy and The Wise Observatory,
	Tel Aviv University, Tel Aviv 69978, Israel
\\	netzer@wise.tau.ac.il
}

\begin{abstract}

We present a series of simulations to
demonstrate that high-fidelity velocity-delay maps
of the emission-line regions in active galactic nuclei
can be obtained from time-resolved spectrophotometric 
data sets like those that will arise from the proposed \Kronos\ satellite.
While previous reverberation-mapping experiments have
established the size scale $R$ of the broad emission-line regions
from the mean time delay $\tau = R/c$ between the
line and continuum variations and have provided
strong evidence for supermassive black holes,
the detailed structure and kinematics of the
broad-line region remain ambiguous and poorly constrained.
Here we outline the technical improvements that will
be required to successfully map broad-line regions
by reverberation techniques.
For typical AGN continuum light curves, characterized by
power-law power spectra $P(f) \propto f^{-\alpha}$
with $\alpha = -1.5\pm0.5$,
our simulations show that a small UV/optical spectrometer
like \Kronos\ will
clearly distinguish between currently viable
alternative kinematic models.
From spectra sampling at time intervals \Tres\ and
sustained for a total duration \Tdur, 
we can reconstruct high-fidelity velocity-delay maps
with velocity resolution comparable to that of the spectra,
and delay resolution $\Delta \tau \approx 2 \Tres$,
provided $\Tdur$ exceeds the BLR crossing time by
at least a factor of three.
Even very complicated kinematical models,
such as a Keplerian flow with superimposed spiral wave pattern,
are cleanly resolved in maps from our simulated \Kronos\ datasets.
Reverberation mapping with \Kronos\ data is therefore likely
deliver the first clear maps of the geometry and kinematics
in the broad emission-line regions
1--100 microarcseconds from supermassive black holes.

\end{abstract}


\section{Introduction}

Reverberation mapping (Blandford \& McKee 1982)
has become part of the standard
toolkit for investigation of active galactic nuclei (AGNs).
Detailed comparison of broad emission-line flux variations
and the continuum variations that drive them
can be used to determine the structure and kinematics of
the broad-line region (BLR) under a rather simple and
straightforward set of assumptions 
(see Horne 1999 and Peterson 2001 for
primers on reverberation mapping).
AGN emission lines are observed to respond roughly linearly
to continuum variations, each line having a different
range of time delays, $\tau$.
Since reprocessing times are relatively short,
the time delays are dominated by light travel time,
\begin{equation}
\label{eq:delay}
	\tau = \frac{R}{c}( 1 + \cos{\theta} )
\ ,
\end{equation}
where ($R$,$\theta$,$\phi$) are spherical polar coordinates
with $\theta=0$ along our line-of-sight beyond
the compact continuum source at $R=0$.
Iso-delay surfaces slice up the geometry of the
emission-line region on a set of nested paraboloids.
In the simplest linear reprocessing model, 
the relationship between the continuum light curve, $F_c(t)$,
and the emission-line profile variations, $F_\ell(v,t)$,
is given by
\begin{equation}
\label{eq:linear}
	F_\ell(v,t) = \int \Psi_\ell(v,\tau) F_c(t - \tau)\, d\tau
\ .
\end{equation}
The ``transfer function,'' or ``velocity-delay map'',
\begin{equation}
\label{eq:psi}
	\Psi_\ell(v,\tau) =
	\frac{1}{d\tau}
	\frac{ \displaystyle \partial F_\ell(v,t) 
	}{ \displaystyle \partial F_c(t - \tau) }
\ ,
\end{equation}
is the response at velocity $v$ and time delay $\tau$
in the flux of emission line $\ell$.
The velocity-delay maps of various emission lines
code information on the geometry, kinematics, and physical
conditions in the BLR.
The immediate goal of reverberation mapping is to recover
$\Psi_\ell(v,\tau)$ for each emission line
by detailed fitting to variability recorded in
high-quality time-resolved spectrophotometric data.
The ultimate goal is to unravel the geometry, kinematics,
and physical conditions in the BLR.

To date, the great success of reverberation mapping has
been determination of the mean broad-line response times
for approximately three dozen AGNs, in several cases for
multiple lines in a single source (see the compilation
of Kaspi et al.\ 2000). By combining the response times
(or ``lags'') with line widths and assuming that gravity
is the dominant force acting on the line-emitting gas,
a virial mass for the central black hole
can be deduced. Two lines of evidence suggest
that these reverberation-based masses are reasonably
accurate and actually do measure the black hole mass:
\begin{enumerate}
\item For AGNs in which the lags and widths of multiple lines have
been measured, there is a clear anticorrelation between
Doppler line width $\vres$ and response time $\tau$
that is consistent with the virial prediction
$\tau \propto \vres^{-2}$
since $\tau$ measures the light-travel time across
the BLR (Peterson \& Wandel 1999, 2000; Onken \& Peterson 2002).
Lines that arise in higher ionization-level
gas (e.g., \heii\,$\lambda1640$, \nv\,$\lambda1240$)
are broader and have shorter response times than
lines that arise primarily in lower ionization-level gas
(e.g., \Hbeta, \ciii]\,$\lambda1909$).
\item Comparison of reverberation-based black-hole masses
and host-galaxy bulge velocity dispersons shows a relationship
that appears to be the same as the black-hole/bulge velocity-dispersion
relationship seen in quiescent galaxies (Ferrarese et al.\ 2001).
\end{enumerate}

On the other hand, reverberation mapping has not yet achieved
the motivating design goal of determining the actual geometry
and velocity field of the BLR.
This is not surprising since inversion of 
eq.\ (\ref{eq:linear})
to solve for the velocity-delay map
requires large amounts of high-quality spectra, which are difficult
to obtain for AGNs. 
Published attempts include analyses
of NGC~4151 (Ulrich \& Horne 1996), and
NGC~5548 (Wanders et al.\ 1995; Done \& Krolik 1996),
both of which yielded rather ambiguous results.
Nevertheless, pursuit of the science goal of determining
the BLR structure is important:
\begin{enumerate}
\item The origin of the BLR emission remains one of the
major mysteries of AGNs. 
There is no widely accepted paradigm for the nature of the BLR. 
The BLR clouds may represent the cooler and denser component
of a two-phase medium in pressure and virial equilibrium
(Krolik, McKee \& Tartar 1981). Alternatively, the BLR clouds
might be in virial motion, but magnetically confined
(Rees 1987).
Compelling arguments have been made
that the broad emission arises as part of a massive outflow 
(e.g., Chiang \& Murray 1996; Bottorff et al.\ 1997). 
Others have suggested that
the broad emission arises in the extended atmospheres
of stars (e.g., Alexander \& Netzer 1997) or 
at least in part from the surface of the accretion disk itself
(e.g., Collin-Souffrin et al.\ 1988).
\item Knowledge of the BLR structure and kinematics is
necessary to understand the potential systematic uncertainties
in reverberation-based black-hole masses. This is especially
interesting if indeed the outflow models are correct since
the virial relationship $\tau \propto \vres^{-2}$ is 
unexpected in simple outflow models. Reverberation-based black-hole
masses are now being used to anchor secondary methods of
mass determination (e.g., Laor 1998; Wandel, Peterson, \& Malkan
1999; Vestergaard 2002) and it is therefore essential to
understand how the measurements are affected by systematics.
\end{enumerate}

Successful reverberation mapping requires a combination of
high time resolution, long duration,
homogenous high signal-to-noise data,
and reasonably high spectral resolution. Precise
specification of these quantities depends on
physical time scales of the source (e.g., BLR light-crossing time)
and the character of the variations, which are neither simple
nor regular. This makes experimental design difficult.
Nevertheless, the first generation of reverberation experiments
have provided a great deal of insight on the 
continuum and emission-line variability properties of AGNs,
making it possible to model their behavior through
numerical simulations. Detailed simulations can
then be used to determine the observational requirements
to map the BLR. 

Our main goal in this paper is to describe
a program we have undertaken to define the requirements
for recovery of high-fidelity velocity-delay maps from data 
obtained with a small UV/optical spectrophotometer.
For the sake of realism, we have used as a model
detector system that of a proposed multiwavelength
observatory, \Kronos\ (Polidan \& Peterson 2001).
The wavelength range and resolution and
achievable signal-to-noise ratios ($S/N$) in these
simulations are set by the specifications for \Kronos.

We have carried out two series of simulations,
a comparatively simple first series,
and a more realistic and complicated second series.
The first series of simulations, discussed in 
\S\ref{sec:series1},
addresses the use of velocity-delay maps for a single line,
for example the \civ\ line, to distinguish among
several alternative and currently viable
models for the geometry and kinematics of the BLR.

The second series, discussed in \S\ref{sec:series2},
is intended to be more realistic and challenging. 
We chose a model with a complicated yet plausible structure,
specifically an inclined Keplerian disk with a 2-armed
spiral density wave superposed on it. 
We then used a photoionization equilibrium model
appropriate for NGC~5548 (Kaspi \& Netzer 1999)
to compute the anisotropic emissivity in each line
at each point in the disk.
In this series of simulations,
we considered the response of multiple lines
and constructed a model spectrum at each time to determine
how blending of closely spaced lines
(e.g., \Lya\,$\lambda1216$ and \nv\,$\lambda1240$)
affect the experimental results.

\section{ Distinguishing Alternative Geometric and Kinematic Models }
\label{sec:series1}

\subsection{Methodology}

Each simulation is based on
theoretical velocity-delay maps for one or more emission lines
corresponding to a particular model for the BLR,
and on fixed observational parameters,
e.g., the sampling rate or time resolution $\Tres$,
the experiment duration $\Tdur$, and the $S/N$ per datum.
The goal of the simulation is to define the extent to which
``observed'' velocity-delay maps, reconstructed from the
simulated data sets, match the theoretical velocity-delay maps.

Each simulation consists of a number of individual realizations,
corresponding to different detailed variations in the driving
continuum light curve and different observational errors.
For each realization, we
(a) generate a model continuum light curve,
(b) produce the corresponding model emission-line light curve
	from the model continuum light curve
	and the theoretical velocity-delay map,
(c) select a simulated data set from pairs of points
	in the model continuum and emission-line light curves,
(d) apply a suitable noise model to the simulated data,
and
(e) analyze the simulated data set by attempting to recover
	the velocity-delay map from it.
Many such realizations are carried out, and the
recovered velocity-delay maps are compared with the
original model to assess the fidelity of the process.
The individual steps in these simulations are
described below.

\subsection{Theoretical Velocity-Delay Maps}

Theoretical velocity-delay maps were constructed for a variety of models. 
In the first series of simulations, the velocity-delay maps were 
selected to be broadly consistent with proposed models for 
\civ\,$\lambda1549$ in NGC~5548
based on the 1993 intensive monitoring program 
with {\em Hubble Space Telescope } (\HST\,), the
{\em International Ultraviolet Explorer} (\IUE\,), and 
optical ground-based telescopes (Korista et al.\ 1995). 
The specific models we consider are:
\begin{enumerate}
\item[A.] A spherical distribution of line-emitting clouds
in circular Keplerian orbits of random inclination,
illuminated by an anisotropic continuum source
(Wanders et al.\ 1995; Goad \& Wanders 1996).
\item[B.] A flat Keplerian disk of clouds.
\item[C.] A hydromagnetically driven wind (Bottorff et al.\ 1997).
\end{enumerate}
The velocity-delay maps for these models are 
presented in Fig.\ \ref{fig:1}.
A common feature of these models is the broad
similarity of their delay maps,
i.e., the line response integrated over Doppler velocity
\begin{equation}
	\Psi_\ell(\tau) = \int \Psi_\ell(v,\tau) dv,
\end{equation}
which underscores the importance of making use of both velocity
and delay information in determining the BLR structure.

\subsection{Model Continuum Variations}

The irregular variations of AGN continua can be
described in terms of
a power-law power-density spectrum
\begin{equation}
P(f) \propto f^{-\alpha}.
\end{equation}
In our first set of simulations, we generate random light
curves with these characteristics by using
the method described by White \& Peterson (1994)
and Peterson et al.\ (1998). For each realization,
a random value of $\alpha$ is selected from
a parent distribution with mean value $\mu(\alpha) = 1.5$
and standard deviation $\sigma(\alpha) = 0.5$
(cf.\ Collier \& Peterson 2001 and references therein).
Each model light curve is then normalized such
that the root-mean-square amplitude of the variations
on a one-month time scale is approximately 16\%.
For the second set of simulations, we used a single
light curve generated from a power spectrum with
$\alpha = 1$, and normalized to a 30\% root-mean-square amplitude
on a 200~day timescale.
In both cases the resulting model continuum light curves
resemble the observed UV continuum variations of well-studied AGNs.

\subsection{Model Line Response}

The line response is computed by
convolving the model continuum light curve
with the model velocity-delay map, as in 
eq.\ (\ref{eq:linear}).
This is straightforward provided the delay interval used to
sample the velocity-delay map is the same as the time interval
\Tres\ of the model light curves.
After the convolution, an appropriate
number of points at the beginning of the light curve is then
removed so that even the first point in the line light curve
is fully responding to the continuum variations.

The velocity sampling $\vres$ used for the velocity-delay maps
should equal or exceed the spectral resolution of the data.
For the first series of simulations, the spectral sampling used
was 2\,\AA\ (i.e, $\sim400$\,\kms\ at \civ\,$\lambda1549$). 
For the second series, we used a spectral sampling roughly
matched to the proposed \Kronos\ spectrographs,
namely 0.7\,\AA\ (1 pixel, $135\,$\kms\ at \civ) 
for the UV lines,
and 2.6\,\AA\ (1 pixel, $160\,$\,\kms\ at \Hbeta). 
for the optical lines. 
The resolution actually achieved in the velocity-delay maps
cannot exceed $\vres$ and $\Tres$, and is limited
by the $S/N$ of the data in comparison with the variations
recorded on different timescales in the light curves.

\subsection{Sampling the Model Data}

For each realization, an artificial data set is 
created by sampling the model data in such a
way as to mimic a real observational program.
Our principal goal is to find the combination
of time resolution $\Tres$ and experiment
duration $\Tdur$ that are required to recover
velocity-delay maps at a high level of confidence,
so most of the simulations considered various
combinations of these two parameters.

In some simulations, we eliminated at random or in groups some  
fraction of the data points to try to realistically 
estimate the effect of data losses.
We found that for the large data sets that we considered,
the results are fairly robust to gaps in the data 
provided a majority of the data points are retained.

\subsection{Simulating Noise}

To simulate observational errors,
each sampled data point is altered by a 
random deviate drawn from a Gaussian distribution
with a mean of zero and appropriate standard deviation.
We assume that the objects to be studied
are the relatively bright AGNs that have already been fairly
well monitored.
For such objects, background noise and readout noise are
insignificant and photon shot-noise is the only
source of statistical uncertainty. 
In the case of a small spacecraft-based observatory
such as \Kronos, the limiting photometric accuracy will probably
be determined by pointing and guiding errors.
We have therefore adopted a very conservative
flux minimum uncertainty of 1\% for each data point.
In the first series of simulations, we assume uncertainties of
$\sim1$\% in continuum and $\sim3$\% in the
emission lines, about the highest precision 
obtained in previous AGN experiments.
In the second series of simulations, the statistical errors are
based on the photon-counting statistics
for a one-hour exposure with \Kronos.

\subsection{Velocity-Delay Maps from the Simulated Data}

The final step in each simulation is to attempt
to recover the velocity-delay map from the
simulated data set. This involves solving
eq.\ (\ref{eq:linear}),
a classical inversion problem in
theoretical physics. In the case of
AGN broad-line reverberation, the data
are relatively sparse and noisy, and usually irregularly sampled:
for this reason, the original Fourier
quotient method proposed by Blandford \& McKee (1982)
has never been applied successfully. 
These complications motivated development of a number of
techniques, superior to Fourier deconvolution, for recovering
the observed transfer function
$\Psi_\ell(v,\tau)$ from relatively poor quality data. These
include the maximum entropy method (MEM) of Horne (1994), the
regularized linear inversion (RLI) method 
of Vio, Horne, \& Wamsteker (1994) and Krolik \& Done (1995), and the
subtractive optimally localized averages (SOLA) method of Pijpers \&
Wanders (1994).

Our simulations utilize the MEM formalism as implememented in 
the software package {\verb+MEMECHO+}, described by Horne (1994). 
We use {\verb+MEMECHO+} to fit the data set, 
defined by the sampled $F_c(t)$ and $F_\ell(v,t)$,
with a linearized echo-mapping model
\begin{equation}
F_\ell(v,t) = \bar{F}_\ell(v) + \int _{0}^{\infty} 
\Psi_\ell(v,\tau) \left[ F_c(t- \tau) - \bar{F}_c\right] \, d \tau
\ .
\end{equation}
This differs from the linear model of eq.\ (\ref{eq:linear})
by the inclusion of ``background'' fluxes
$\bar{F}_c$ for the continuum,
and $\bar{F}_\ell(v)$ for the line profile.
These backgrounds account for approximately time-independent
contaminating fluxes,
including the host-galaxy starlight contribution to the continuum 
and narrow emission-line flux in the core of the emission line
profile.
We set the backgrounds to the mean line and continuum levels in the data.
The velocity-delay map $\Psi_\ell(v,\tau)$ 
in the above model then represents the marginal
responsivity (i.e., relative to the background levels)
to changes in the continuum. 
The linearized echo-mapping model is thus insensitive
to mild non-linear line responsivities
and affords an improvement over the simpler linear model of 
eq.\ (\ref{eq:linear}).
The MEM fit proceeds iteratively, by varying model $F_c(t)$, 
$\Psi_\ell(v,\tau)$,
and $\bar{F}_\ell(v)$ to fit simultaneously the observed $F_c(t)$ and
$F_\ell(v,t)$. 
The ``smoothest positive maps'', i.e., 
the maps which maximize the entropy,
for which $\chi^{2} /N \approx 1 \pm \sqrt{2/N}$,
with $N$ the number of line and continuum measurements,
defines a satisfactory fit to the data. 
The ``observed'' velocity-delay map
$\Psi_\ell(v,\tau)$ that is recovered from the simulated
data is then compared with the original model velocity-delay map.

\subsection{ Discussion of Results }

Our first series of simulations has been described in some
detail in a conference proceedings (Collier, Peterson, \& Horne 2001),
so here we report on some highlights that
illustrate our general conclusions.

Our first general finding supports the ``rule of thumb''
(e.g., Press 1978)
that the time series should be at least three times longer
than the maximum time scale to be probed. The BLR models
used here have an outer radius $\Rout \approx 10$ light days
(varying slightly for the particular models)
so the longest timescale for response is $2\Rout/c \approx
20$\,days. The velocity-delay maps recovered from the simulated
data became recognizable and distinguishable from one another only
with $\gtsim60$ days of data, regardless of how fine a sampling
grid is used. 

In Fig.\ \ref{fig:2}, we show the ``observed'' velocity-delay maps
recovered from a simulation with duration $\Tdur = 60$\,days
and time resolution $\Tres = 0.1$\,day
(i.e., at total of $\sim600$ observations).
In comparison with the corresponding theoretical maps in Fig.\ \ref{fig:1},
these ``observed'' maps are clearly somewhat blurred
in both the delay and velocity dimensions.
The finite resolution of the reconstructed maps is a limiting
factor in their ability to distinguish among alternative
geometric and kinematic models of the BLR.
High resolution maps require good sampling in time and velocity,
and a sufficient signal-to-noise in the observations to detect
the possibly quite small brightness changes that provide the
information that defines the maps.
Comparison of Figs.\ref{fig:1} and \ref{fig:2} indicates that
the resolution achieved in Fig.\ \ref{fig:2} 
is about $1$ day in $\tau$ and a few hundred 
km~s$^{-1}$ in $v$.
We consider this to be adequate
to distinguish among the three models shown.

For somewhat coarser time resolution, longer durations are required
to achieve comparable fidelity, though fewer data points suffice:
for results similar in quality to those shown in Fig.\ \ref{fig:2},
$\Tres = 0.5$\,day requires $\Tdur \approx 80$\,days ($\sim400$ observations)
and
$\Tres = 1$\,day requires $\Tdur \approx 100$\,days ($\sim100$ observations).
However, further increases in $\Tres$ resulted in unacceptable
loss of delay resolution in the velocity-delay map; in these simulations
the shortest time scales of interest were of order 2 days, so the
resolution of the monitoring experiment must be finer than this.

At first glance, it appears that these requirements have
been {\em nearly} met in previous reverberation-mapping
programs, most notably the 1996 \IUE\
campaign on NGC~7469 (Wanders et al.\ 1997), which arguably represents
the current state of the art. 
We test the verisimilitude
of our simulations by reproducing the characteristics of
this previous observing campaign.

We therefore carried out the same simulation process,
with minor changes made to mimic the characteristics
of the \IUE\ data. Specifically, we adopted the 
following parameters:
(a) $\Tdur=49$\,days, (b) $\Tres=0.2$\,days, 
(c) continuum and emission-line flux uncertainties of 3\% and 7\%, 
respectively, 
(d) spectral resolution $\sim8$\,\AA, and 
(e) continuum variations normalized to an amplitude of
12\% on timescales of 10 days. 
We show the results of these simulations for two of our model
velocity-delay maps 
in Fig.\ \ref{fig:3},
which should be compared with the corresponding
models in Fig.\ \ref{fig:1}.
The delay resolution of these maps is now $\sim5$ days,
owing mainly to the lower $S/N$ compared with the maps in
Fig.\ \ref{fig:2}.
The quality of the velocity-delay maps
that we obtain in these simulations is similar to
the quality of the velocity-delay maps recovered from
the real data.
It is clear that the recovered velocity-delay maps
based on simulations using \IUE-like parameters do not have
sufficient $S/N$ and/or duration to distinguish between
these two competing BLR models. 

\section{Simulations of a Photoionized Disk with Spiral Arms}
\label{sec:series2}

The second series of simulations is intended to be more
challenging: the simulated data are more realistic and
the BLR structure is chosen to be rather complicated.
The precise nature of the BLR structure is less
important than the fact that it is complex: the 
desired demonstration is that a complex system can be
successfully mapped.

\subsection{Spiral Disk Kinematics}

In these simulations, the BLR model we adopted
is a population of $10^5$ 
discrete gas clouds on elliptical Keplerian orbits
around a black hole of mass $3\times10^7$\,\Msun.
The orbital semi-major axes $a$ are populated with gas clouds
uniformly distributed in $\log{a}$ from 
$a_{\rm in}=3$ to $a_{\rm out}=50$ light days.
The orbits all lie in a plane inclined at $i=45\deg$ to the line of 
sight. 

To generate a spiral wave pattern on the Keplerian disk,
we make the orbits elliptical, in this case eccentricity $e=0.3$,
and we ``twist'' the orbits by advancing the azimuth $\theta$ of
the major axis with $\log{a}$:
\begin{equation}
	\theta_0(a) = \theta_{\rm in}
	+ \ln{\left( \frac{ a_{\rm in} - a }{ \Delta a} \right)}
\ .
\end{equation}
The orbiting gas clouds spend most of their time near
the apbothron, at $(R,\theta) = \left( a/(1-e), \theta_0(a) \right)$.
The twisting set of elliptical orbits thus generates a 
trailing spiral wave pattern,
with the surface density of clouds on the face of the disk
enhanced near
\begin{equation}
	R(\theta) = \frac{a_{\rm in}
	- \Delta a\
	\exp{ \left\{ \theta - \theta_{\rm in} \right\} }
	}{ 1 - e }
\ .
\end{equation}
To produce two such spiral arms, we simply rotate the orientation
of the orbit of each cloud by $180^\circ$ with 50\% probability.

\subsection{ Physical Conditions in the Model BLR }

Physical conditions in the clouds vary with their radial distance
$R$ from the nucleus.
We adopt a power-law model similar to that which
Kaspi \& Netzer (1999) employed in their attempt to fit
the observed behavior of NGC~5548 during the
1989 monitoring campaign by the International AGN Watch
(Clavel et al.\ 1991; Peterson et al.\ 1991).
The gas density is uniform inside a given cloud,
and differs from cloud to cloud,
decreasing outward as
\begin{equation}
	n = n_1 \left( \frac{R}{R_1}\right)^{-s}
\ ,
\end{equation}
where we adopt $s=1$ for the density exponent
and $n_1=10^{11.4}\cc$ for the density
at the fiducial radius $R_1=10^{15.4}$cm$=1$ light day.
Note that $R$ varies from $a/(1+e)$ to $a/(1-e)$,
and the density of a given cloud varies accordingly,
as the cloud traverses its elliptical orbit.
The column density is similarly given by
\begin{equation}
	N = N_1 \left( \frac{R}{R_1}\right)^{-2s/3}
\ ,
\end{equation}
with $N_1=10^{24}$cm$^{-2}$, and
the corresponding ionization parameter is
\begin{equation}
U \equiv \frac{Q}{ 4\pi c R^2 n}
	= U_1 \left( \frac{R}{R_1}\right)^{s-2}
\ ,
\end{equation}
where $U_1\approx10^{0.2}$
for the adopted mean ionizing photon rate,
$\bar{Q}=10^{54}$~photons~s$^{-1}$.
For $3 < R/R_1 < 50$, we have 
$10.92 \geq \log{n} \geq 9.40$,
$23.68 \geq \log{N} \geq 22.67$, and
$-0.28 \geq \log{U} \geq -1.80$.
This model roughly reproduces the emission line
ratios observed in NGC~5548.

Individual clouds have a projected area 
$A \propto (N/n)^2 \propto R^{2s/3}$,
and therefore cover a solid angle
$\Omega = A/R^2 
\propto R^{\frac{2s}{3}-2} \propto R^{-4/3}$
as viewed from the nucleus.
We normalize the cloud population by
summing over all clouds to obtain a total solid angle
$\sum_i \Omega_i =4\pi f$,
and set the total covering fraction to $f=0.3$.
This simply scales the model line fluxes to
roughly reproduce those observed
in NGC~5548 (Kaspi \& Netzer 1999).

Our assumed thin disk geometry is probably inconsistent
with so large a covering fraction as $f=0.3$, unless
anisotropic emission or scattering of the ionizing radiation
increases its probability of crossing the disk plane.
Obscuration of one cloud by another is also ignored.
We largely overlook these issues here, because our
main purpose is not to produce the most realistic BLR model,
but rather to check the ability of \Kronos\ to recover rich 
structure in the velocity-delay map of a complicated BLR.

\subsection{ Reprocessing by the Photoionized Gas Clouds }

We employ the photoionization model {\verb+ION+}
(Kaspi \& Netzer 1999 and references therein)
to compute inward and outward emission-line fluxes, 
$F^-_\ell(n,N,U)$ and $F^+_\ell(n,N,U)$,
as functions of 
density $n$, column density $N$, and ionization parameter $U$.
For intermediate viewing angles we interpolate 
linearly in $\cos{\theta}$:
\begin{equation}
F_\ell(\theta) = \left( \frac{F^+_\ell + F^-_\ell}{2} \right)
	+  \left( \frac{F^+_\ell - F^-_\ell}{2} \right)
	\cos{\theta} 
\ .
\end{equation}
To save computer time, we pre-compute the inward and outward
fluxes for each line on a 3-dimensional grid spanning
$7.5 \leq \log{n} \leq 13$,
$19 \leq \log{N} \leq 23.4$,
and
$-3.5 \leq \log{U} \leq 0.25$,
at intervals
$\Delta\log{n}=0.5$,
$\Delta\log{N}=0.4$,
and 
$\Delta\log{U}=0.25$,
respectively.
We interpolate in this grid to compute
line fluxes contributed by each cloud at each location in its
orbit in each of five key emission lines,
\Lya, \nv\,$\lambda1240$,
\civ\,$\lambda1549$, \heii\,$\lambda1640$, 
and \Hbeta.

\subsection{ Theroetical Velocity-Delay Maps }

Fig.\ \ref{fig:4} shows a color representation of
the spiral disk BLR model projected on the sky,
and the corresponding velocity-delay map.
To generate these color representations we
superimpose the responses of three different emission lines,
\Lya, \civ, and \heii,
in red, blue, and green, respectively.
The two spiral arms clearly visible on the sky view
are easily traced on the velocity-delay map.
The vivid colors apparent in both the sky view and the
velocity-delay map arise because each line
probes a different range of physical conditions
and ionization level.
The \heii\ zone, for example, is more tightly confined
to the vicinity of the nucleus.
The front-to-back asymmetry reflects different
anisotropies in the emission of different lines;
the \Lya\ emission, in particular, is stronger
on the far side of the nucleus.

\subsection{ Ionizing and Continuum Light Curves }

Variations in the ionizing radiation are generated by using
a random walk in time, which has 
a power spectrum $P(f) \propto f^{-1}$.
We first standardize the random walk by
subtracting its mean and dividing by its standard deviation,
evaluated at the times of observations,
to obtain $\eta(t)$ with mean 0 and standard deviation 1.
We can then scale $\eta(t)$ to match the desired mean  $\bar{Q}$
and amplitude  $\Delta Q$ of the variable ionizing radiation.
To ensure that $Q(t)$ can never become negative, we take
\begin{equation}
	Q(t) = \bar{Q}\ \exp{ \left\{ \Delta Q\ \eta(t) / \bar{Q} 
			\right\} }
	\approx  \bar{Q} + \Delta Q\ \eta(t)
\ .
\end{equation}
For these simulations we take
$\bar{Q}=10^{54}$ photons~s$^{-1}$ for the mean ionizing photon rate,
and $\Delta Q=0.3 \bar{Q}$
for the root-mean-square amplitude of the variations in ionizing flux
over the 200-day period.

We assume that the UV and optical continuum variations $F_c(\lambda,t)$
are proportional the ionizing photon variations $Q(t)$,
\begin{equation}
        F_c(\lambda,t) = \bar{F}_c(\lambda)\
	\exp{ \left\{ \Delta F_c(\lambda)\ \eta(t) / \bar{F}_c(\lambda) 
			\right\} }
        \approx  \bar{F}_c(\lambda) + \Delta F_c(\lambda)\ \eta(t)
\ .
\end{equation}
Here $\bar{F}_c(\lambda)$ and  $\Delta F_c(\lambda)$ 
are the mean and root-mean-square continuum spectra,
for which we adopt power-law forms matched to the
observed mean and root-mean-square variations 
in the continuum of NGC~5548 at $135$~nm and $510$~nm.
Note that the UV and optical continua may arise in part from
reprocessing of harder photons, and hence may exhibit a
wavelength-dependent range of time delays.
For the present simulations we neglect this possibility,
since the continuum time delays are much smaller than those
in the emission lines.

\subsection{ Synthetic \bfKronos\ Spectra }

In addition to using the photoionization model 
to account for
the non-linear and anisotropic responses of each cloud to
the ionizing radiation, Doppler shifts and time delays are taken into
account when adding to the spectrum at each time the line emission
contributions from each gas cloud:
\begin{equation}
	F_\ell( v , t ) =
	\sum_{i}
	\frac{ A_i F_\ell\left( \theta_i, n_i, N_i, U_i(t) \right)
		}{ D_L^2(z) }
	\frac{ \displaystyle
		\exp{ \left\{ - \frac{1}{2} 
	\left( \frac{ v - v_i }{\vres_i} \right)^2 \right\} }
	}{
		 \sqrt{2 \pi} \vres_i
	}
\ ,
\end{equation}
where $z$ is the redshift, and $D_L(z)$ the luminosity distance.
At time $t$,
cloud $i$, with density $n_i$, column density $N_i$ and area $A_i$,
is located at $R_i$, $\theta_i$, and
has an ionization parameter $U_i(t) = Q(t-\tau_i)/4\pi c R_i^2 n_i$,
where the light travel time delay is
$\tau_i = (R_i/c)(1+\cos{\theta_i})$.
In addition to the orbital Doppler shift, $v_i$,
each cloud is given a velocity dispersion 
$\vres_i = 0.1$ times
the local Keplerian orbital velocity.

Synthetic spectra are generated using the design spectral
range, resolution, and wavelength-dependence sensitivity
of the \Kronos\ spectrometers (Polidan \& Peterson 2001).
Specifically, we assume UV and NUV/Opt spectrographs covering
100--175~nm with 1100 pixels and 270--540~nm with 2000 pixels,
respectively.
The noise model adopted in the simulations
is based on photon-counting statistics,
assuming an exposure time of 3600~s,
and wavelength-dependent effective area
$A_{\rm eff}(\lambda)$ interpolated in wavelength
from the values given in Table~1.

\begin{table}
\begin{center}
{\bf Table 1. \bfKronos\ Spectrographs }
\\
\begin{tabular}{ccc|ccc}
\hline
$\lambda$ & $A_{\rm eff}$ & \vres &
$\lambda$ & $A_{\rm eff}$ & \vres
\\ (nm) & (cm$^2$) & (km~s$^{-1}$pix$^{-1})$
& (nm) & (cm$^2$) & (km~s$^{-1}$pix$^{-1})$
\\ \hline \hline
   100 & 232 & 225 & 270 & 584 & 150
\\ 112 & 299 & 200 & 300 & 623 & 135
\\ 125 & 259 & 180 & 350 & 600 & 116
\\ 150 & 154 & 150 & 400 & 675 & 101
\\ 160 & 112 & 140 & 450 & 668 & 90
\\ 175 &  72 & 129 & 500 & 567 & 81
\\     &     &     & 540 & 510 & 75
\\ \hline
\end{tabular}
\end{center}
\end{table}

The quality of AGN spectra that will be obtained by \Kronos\
in a 1-hour exposure on NGC~5548 is indicated in Fig.\ \ref{fig:5}.
Here the top spectrum is our synthetic spectrum
for a single 1-hour integration obtained near peak brightness.
Below this (with much less noise) are the mean and root-mean-square
of 1001 such synthetic spectra, from a simulation with
$\Tres = 0.2$\,day and $\Tdur=200$\,days.

\subsection{ Recovery of Delay Maps }

To extract continuum and emission-line light curves from
the synthetic spectra, we employ
the same code that we use to measure real spectra.  
A power-law continuum is fitted to each synthetic 
spectrum, and light curves for the continuum and
continuum-subtracted line fluxes
are measured from the spectra.
{\verb+MEMECHO+} is then used to fit the
measured UV continuum light curve and the continuum-subtracted
light curves of each emission line.
As shown in Fig.\ \ref{fig:6},
this fit reproduces the observed continuum
variations and the variously time-delayed line flux variations,
recovering a delay map $\Psi_\ell(\tau)$ for each line.

\subsection{ Recovery of Velocity-Delay Maps }

Finally, the continuum-subtracted line variations
are measured for typically 100 bins 
across the velocity profile of each emission line.
A {\verb+MEMECHO+} fit then recovers
a velocity-delay map $\Psi_\ell(v,\tau)$ for each line. 
The results are shown in Fig.\ \ref{fig:7}.
The spiral patterns are clearly visible in both the sky views
and the theoretical velocity-delay maps.
The reconstructed maps are again blurred in $\tau$ and $v$,
due to the finite resolution of the mapping procedure,
but the spirals are clearly recovered for the strong
\Lya, \civ, and \Hbeta\ lines, and incipient structure is
apparent even in the weaker, rapidly responding \heii\ line.

As with the first set of simulations, we find that 
when we increase $\Tres$, we must compensate by
increasing $\Tdur$ to recover
velocity-delay maps of comparable fidelity.
This enables mapping over time delays longer than $\sim 2\Tres$
with a smaller total number of spectra.
However, the structure lost on short time scales
leads to highly indeterminate results
for rapidly responding lines like \heii.
Furthermore, there are few AGNs that can be
observed continuously for periods longer than $\sim200$\,days
with an Earth-orbiting observatory without the target
getting too close to the Sun for observation.

\section{Discussion}

The major result of this investigation is  a clear
demonstration that with technically realizable observational programs
reverberation mapping can successfully recover even
complex emission-line velocity-delay maps.

Previous reverberation mapping programs have had comparatively
modest goals: generally, the intent has been to measure
the mean response time for various emission lines.
From these programs, we have 
learned enough about AGN continuum and emission-line variability
characteristics to carry out realistic simulations, such as
those described here, that will define future programs with
more ambitious goals. These simulations show clearly that 
even the most ambitious previous programs could not be
expected to yield the results we now seek, i.e., a complete
velocity-delay map that can be used to identify the
detailed structure and kinematics of the BLR.

We also conclude from these simulations that while 
high-fidelity reverberation mapping of
even a single line will be a tremendous step forward, 
this will not yield the complete structure of the BLR. 
This is clearly illustrated in Fig.\ \ref{fig:4},
where the vivid colors arise because
different emission lines probe
different ranges of physical conditions
and ionization level.
To acquire a complete picture of the BLR, a variety of lines spanning 
a broad range of ionization level (and hence mean response time)
need to be mapped.
Moreover, it is distinctly possible that the high-ionization and 
low-ionization lines arise primarily in physically distinct regions with
different geometries and kinematics (e.g., Collin-Souffrin
et al.\ 1988); if this is true, the need
for reverberation mapping of multiple emission lines is self-evident.

Once we have acquired the data to make high-fidelity velocity-delay maps
for different emission lines, how can we produce a map of the BLR? 
Perhaps the simplest approach is though inspired modeling:
practitioners recognizing the structure in the velocity-delay maps
can devise appropriate models with adjustable parameters
to obtain the best fit of a model to the data. 
A more ambitious program aims to reconstruct a complete
phase-space map from the velocity-delay maps of different lines, 
or even directly from the data.
This may be problematic, however, because the ``observable''
velocity-delay map is a two-dimensional projection of
the six-dimensional phase space, so there are degeneracies. 
These can be partially resolved by combining results from
multiple lines with photoionization equilibrium models,
and some success has already been achieved with simple
geometries (Horne 2001, Horne, Korista, \& Goad 2002).
Such methods will succeed best if the BLR structure
has some degree of simplifying symmetry,
as we expect from most current models.
But even if the BLR is completely chaotic
with no symmetries,
this could be concluded from good reverberation data,
and would provide us with an important answer
about the inner structure of AGNs.

\acknowledgements{We are grateful to NASA for support
of this work through LTSA Grant NAG5--8397 to
The Ohio State University.
KH was supported by a PPARC Senior Fellowship,
and by a Beatrice Tinsley Visiting Professorship
at the University of Texas at Austin. We thank Mark
Bottorff for supplying the model transfer function 
shown in Fig.\ 1c.
}

\bigskip

\clearpage

\begin{figure}
\caption[]{
Theoretical velocity-delay maps $\Psi_\ell(v,\tau)$ for
(a) a spherical distribution of line-emitting clouds
in circular Keplerian orbits of random inclination,
illuminated by an anisotropic continuum source
(Wanders et al.\ 1995; Goad \& Wanders 1996),
(b) a flat Keplerian disk of clouds, and
(c) a hydromagnetically driven wind (Bottorff et al.\ 1997).
Projections $\Psi_\ell(v)$ and  $\Psi_\ell(\tau)$
are shown below and to the right, respectively, of
each greyscale $\Psi_\ell(v,\tau)$ map.
}
\label{fig:1}
\end{figure}

\begin{figure}
\caption[]{
Observed velocity-delay maps $\Psi_\ell(v,\tau)$
recovered from a simulation with duration $\Tdur = 60$\,days
and time resolution $\Tres = 0.1$\,day (i.e., at total of $\sim600$ 
observations).
As in Fig.\ \ref{fig:1}, the three cases shown are for
(a) a spherical distribution of line-emitting clouds
in circular Keplerian orbits of random inclination,
illuminated by an anisotropic continuum source
(Wanders et al.\ 1995; Goad \& Wanders 1996),
(b) a flat Keplerian disk of clouds, and
c) a hydromagnetically driven wind (Bottorff et al.\ 1997).
Projections $\Psi_\ell(v)$ and  $\Psi_\ell(\tau)$,
with Monte Carlo error bars,
are shown below and to the right, respectively, of
each greysacle $\Psi_\ell(v,\tau)$ map.
}
\label{fig:2}
\end{figure}

\begin{figure}
\caption[]{
Observed velocity-delay maps $\Psi_\ell(v,\tau)$ recovered from 
simulated data with characteristics similar
to continuous \IUE\ monitoring,
specifically, $\Tdur=49$\,days, $\Tres =0.2$\,days, 
continuum and emission-line flux uncertainties of 3\% and 7\%, 
respectively, 
spectral resolution $\sim8$\,\AA, and 
continuum variations normalized to an amplitude of
12\% on timescales of 10 days. 
The two cases shown, as in Fig.\ \ref{fig:1}, are 
(a) a spherical distribution of line-emitting clouds
in circular Keplerian orbits of random inclination,
illuminated by an anisotropic continuum source
(Wanders et al.\ 1995; Goad \& Wanders 1996), and
(b) a flat Keplerian disk of clouds.
Projections $\Psi_\ell(v)$ and  $\Psi_\ell(\tau)$,
with Monte Carlo error bars,
are shown below and to the right, respectively, of
each greysacle $\Psi_\ell(v,\tau)$ map.
}
\label{fig:3}
\end{figure}

\begin{figure}
\caption[]{
Velocity-delay map and corresponding sky view
for a Keplerian accretion disk with two spiral arms.
Red, green, and blue correspond to 
the \Lya, \civ, and \heii\ emission lines, respectively.
}
\label{fig:4}
\end{figure}

\begin{figure}
\caption[]{
Synthetic \Kronos\ spectra for NGC~5548.
Top spectrum is a single 1-hour integration near peak brightness,
below which are the
mean and root-mean-square of 1001 such spectra,
representing 1-hour integrations every 0.2 days for 200 days.
This simulation of reverberation in the environment
of a $10^7M_\odot$ black hole includes a Keplerian disk
with spiral density waves,
continuum variations, and responses in five key emission lines,
\Lya, \nv\,$\lambda1240$,
\civ\,$\lambda1549$, \heii\,$\lambda1640$, 
and \Hbeta.
}
\label{fig:5}
\end{figure}

\begin{figure}
\caption[]{
Continuum and velocity-integrated emission line fluxes
measured from  simulated \Kronos\ spectra,
representing 1-hour integrations every 0.2 days for 200 days.
The driving continuum light curve $F_c(t)$ is in the bottom panel,
the responding line light curves $F_\ell(t)$ above,
and the recovered delay maps $\Psi_\ell(\tau)$ 
are shown in the left-hand column.
}
\label{fig:6}
\end{figure}

\begin{figure}
\caption[]{
Recovery of velocity-delay maps from simulated \Kronos\ data sets.
The left column shows ``sky views'' of the
adopted BLR model, in this case a 
Keplerian accretion disk with two spiral arms,
in the light of four different emission lines.
The middle column shows the corresponding 
theoretical velocity-delay maps,
in which the spiral arms are also clearly visible.
The right column gives the ``observed'' velocity-delay maps
recovered from simulated \Kronos\ data
representing 1-hour exposures every 0.2 days for 200 days.
}
\label{fig:7}
\end{figure}

\clearpage

\end{document}